# Generalized measurement of uncertainty and the maximizable entropy

Congjie Ou[1,2], Aziz El Kaabouchi[1], Alain Le Méhauté[1], Qiuping A. Wang[1], Jincan Chen[1,3]

[1]Institut Supérieur des Matériaux et Mécaniques Avancés du Mans
44, Avenue F.A. Bartholdi, 72000 Le Mans, France

[2]College of Information Science and Engineering, Huaqiao University,
Quanzhou 362021, China

[3]Department of Physics and Institute of Theoretical Physics and Astrophysics
Xiamen University, Xiamen 361005, China

**Abstract:** For a random variable $x$ we can define a variational relationship with practical physical meaning as $dI = d\bar{x} - \overline{dx}$, where $I$ is called as uncertainty measurement. With the help of a generalized definition of expectation, $\bar{x} = \sum_i g(p_i) x_i$, and the expression of $dI$, we can find the concrete forms of the maximizable entropies for any given probability distribution function, where $g(p_i)$ may have different forms for different statistics which includes the extensive and nonextensive statistics.

# 1. Introduction

For a random dynamical system, the probability distribution function is very important to describe the observable quantities of the system and the other corresponding characteristics. A physical system may have many different microstates and each of them has a probability. They are indeterminate unless the corresponding probability equal to 0 or 1. Therefore, there exists a measurement of uncertainty, or sometimes called information, for such a system. It's well known that the Shannon informational entropy [1], i.e., $S = -\sum_i p_i \ln p_i$, has been widely used for dynamical systems in equilibrium such as the Boltzmann-Gibbs statistics [2]-[4] and nonequilibrium [5][6]. Recently, accompany with more and more peculiar experimental phenomena [7][8], nonextensive statistical mechanics (NSM) was proposed [9]-[12] to describe such complex systems presenting long-range interactions and/or long-duration memory. The fundamental quantity defined in NSM is a nonextensive entropy with an additional nonextensive parameter. When the parameter tends to unity all the results will reduce to those obtained in the Boltzmann-Gibbs statistics. The nonextensive entropy can yield different kinds of observed peculiar distribution functions via the maximum entropy method (MaxEnt). If we consider MaxEnt as a basic principle for an arbitrary dynamical system, there should exist one corresponding maximizable entropy form for each observed peculiar probability distribution. How to search a proper entropy form for a given probability distribution function? This question has recently investigated by Wang [13] and some important entropy forms have been reconstructed with the help of a variational definition. In the present paper, we go further on this topic by using a generalized definition of expectation value. In section 2 a generalized varentropy definition is given. Then some useful entropic forms are derived in section 3. In section 4 we get some novel and general results not only suitable for extensive statistics but also for the nonextensive one.

## 2. A generalized definition of varentropy

The variational uncertainty measurement for a random variable $x$ can be written as

$$dI = d\bar{x} - \overline{dx}. \tag{1}$$

where " $\bar{\ }$ " means the expectation or average value of a quantity. For different statistics, there exist different kinds of expectation definition such as the traditional expectation $\bar{x} = \sum_i p_i x_i$, incomplete expectation $\bar{x} = \sum_i p_i^q x_i$, etc., where $x_i$ is the $i$th possible value of $x$ and $p_i$ is the corresponding probability. So we propose a unified method to represent all these different expectation definitions, which reads

$$\bar{x} = \sum_i g(p_i) x_i, \tag{2}$$

where $g(p_i)$ is a function of probability distribution $\{p_i\}$ $(i = 1,2,...)$. Substituting Eq. (2) into Eq. (1), one can easily get

$$dI = \sum_i x_i dg(p_i). \tag{3}$$

In fact Eq. (2) implies that in the case of $\bar{x} = x_i = x_j = const$, the probability normalization condition is given by

$$\sum_i g(p_i) = 1. \tag{4}$$

The derivative of Eq. (4) with respect to $p_i$ gives

$$\sum_i dg(p_i) = 0. \tag{5}$$

It's worth to point out that Eq. (1) from Ref. [13] can be considered as the measure of probabilistic uncertainty for the random variable $x$. As long as the probability distribution of a system is given, one can derive a unique entropic form for such a system according to the definition of expectation resulting from the concrete statistical method. Furthermore, this entropy is maximizable by the Lagrange multiplier method and finally yields the same distribution function.

## 3. Applications: reconstruct some interest entropy forms

In this section, we choose some well-known probability distribution functions to derive the corresponding entropy forms. For example, let's consider the power law distribution $p_i = e^{-\beta x_i}/Z$ where $\beta$ is a positive constant and $Z$ is the partition function. It's one of the most frequently used distribution forms and the corresponding expectation reads $\bar{x} = \sum_i p_i x_i$, which means $g(p_i) = p_i$. So from Eqs. (3)-(5) one can easily get

$$dI = \sum_i x_i dp_i = \sum_i \frac{-1}{\beta} \ln(p_i Z) dp_i, \qquad (6)$$

Integrating Eq. (6) yields

$$I = \frac{-1}{\beta} \sum_i p_i \ln p_i + C. \qquad (7)$$

Comparing the Shannon entropy $S = -\sum_i p_i \ln(p_i)$ with Eq. (7) one can easily find that $I = S/\beta$. For the sake of convenience we can set $\beta = 1$ below, and the integral constant $C$ should be equal to zero if we substitute $I(p_i = \delta_{1,i}) = 0$ into Eq. (7). This example is the same with that obtained in Ref. [13] because in this case $g(p_i) = p_i$ is chosen. Under this simple probability normalization one can also derive the stretched exponential entropy, Kappa entropy and the maximizable entropy for Cauchy distribution [15]. In the present paper we introduce Eq. (3) as a generalization of the varentropy definition in Ref. [13]. For any observed probability distribution $\{p_i(x_i)\}$, if one choose a kind of statistics, i.e., $g(\{p_i(x_i)\})$, to describe it, there exists a corresponding entropy form.

If $p_i = \frac{1}{Z}\left[1-(1-q)\frac{x_i - \bar{x}}{\sum_j p_j^q}\right]^{\frac{1}{1-q}}$ is taken into account [10], it means that $g(p_i) = \frac{p_i^q}{\sum_j p_j^q}$,

where $\bar{x} = \frac{\sum_i p_i^q x_i}{\sum_j p_j^q}$. So one can write

$$dg(p_i) = \sum_k \left[\frac{qp_i^{q-1}\delta_{i,k}dp_k}{\sum_j p_j^q} - \frac{p_i^q \sum_j qp_j^{q-1}\delta_{j,k}dp_k}{\left(\sum_j p_j^q\right)^2}\right]$$
$$= \frac{qp_i^{q-1}dp_i}{\sum_j p_j^q} - \frac{p_i^q \sum_j qp_j^{q-1}dp_j}{\left(\sum_j p_j^q\right)^2}$$
(8)

Substituting Eq. (8) into Eq. (3) one can get

$$dI = \frac{\sum_i qp_i^{q-1}x_i dp_i}{\sum_j p_j^q} - \frac{\sum_i p_i^q x_i \sum_j qp_j^{q-1}dp_j}{\left(\sum_j p_j^q\right)^2}$$
$$= \frac{\sum_i qp_i^{q-1}x_i dp_i}{\sum_j p_j^q} - \bar{x}\frac{\sum_j qp_j^{q-1}dp_j}{\sum_j p_j^q}$$
(9)

On the other hand, from the distribution function one can get

$$x_i = \frac{1-(p_i Z)^{1-q}}{(1-q)}\sum_j p_j^q + \bar{x}.$$
(10)

Substituting Eq. (10) into (9) yields

$$dI = \frac{\sum_i qp_i^{q-1}\frac{1-(p_i Z)^{1-q}}{(1-q)}\sum_j p_j^q dp_i}{\sum_j p_j^q} + \bar{x}\frac{\sum_i qp_i^{q-1}dp_i}{\sum_j p_j^q} - \bar{x}\frac{\sum_j qp_j^{q-1}dp_j}{\sum_j p_j^q}$$
$$= \sum_i qp_i^{q-1}\frac{1-(p_i Z)^{1-q}}{(1-q)}dp_i$$
$$= \frac{q}{(1-q)}\sum_i p_i^{q-1}dp_i$$
(11)

then

$$I = \frac{\sum_i p_i^q}{1-q} + C. \tag{12}$$

If one can choose $I = 0$ when $p_i = \delta_{1,i}$ the integer constant $C$ is determined by

$$C = \frac{-1}{1-q}, \tag{13}$$

And consequently,

$$I = \frac{\sum_i p_i^q - 1}{1-q}. \tag{14}$$

Eq. (14) is nothing but the Tsallis entropy [9][10].

If $p_i = \frac{1}{Z}\left[1-(1-q)q\frac{x_i - \bar{x}}{\sum_j p_j}\right]^{\frac{1}{1-q}}$ is taken into account [14], it means that $g(p_i) = p_i^q$,

where $\bar{x} = \sum_i p_i^q x_i$. From this distribution function one can get

$$x_i = \frac{1-(p_i Z)^{1-q}}{(1-q)q}\sum_j p_j + \bar{x}. \tag{15}$$

Substituting $g(p_i)$ and Eq. (15) into Eq. (3) yields

$$dI = \sum_i \frac{1-(p_i Z)^{1-q}}{(1-q)q}\left(\sum_j p_j\right)qp_i^{q-1}dp_i + \bar{x}\sum_i qp_i^{q-1}dp_i. \tag{16}$$

It can be seen from Eq. (5) that the second item on the right hand side of Eq. (16) is equal to zero. On the other hand, to rewrite the incomplete probability distribution yields

$$(p_i Z)^{1-q} = 1-(1-q)q\frac{x_i - \bar{x}}{\sum_j p_j}. \tag{17}$$

Multiplying $p_i^q$ to the both sides of Eq. (17) and summing for all $i$ we can have

$$\sum_j p_j = Z^{q-1}. \tag{18}$$

From Eqs. (16) and (18) we can derive

$$dI = \frac{-\sum_i dp_i}{1-q}, \qquad (19)$$

and then

$$I = \frac{-\sum_i p_i + C}{1-q}. \qquad (20)$$

By the same trick of the above one we can determine the constant $C = 1$ and finally we get

$$I = \frac{1 - \sum_i p_i}{1-q}. \qquad (21)$$

It's just the entropic form of incomplete statistics [11].

In fact, the non-power law distribution has been observed in more and more physical systems [7][8][16]-[19] and it is named by $q$-power law. The distribution can be written as $p_i(x_i) = \frac{1}{Z}[1-(1-q)x_i]^{1/(1-q)}$, which presents different characteristics with different q values. According to Eq. (3) we can also find a maximizable entropy for this distribution. From

$$x_i = \frac{1-(p_i Z)^{1-q}}{1-q} \qquad (22)$$

and $g(p_i) = p_i$ one can get

$$I = \frac{Z^{1-q}(1-\sum_i p_i^{2-q})}{(1-q)(2-q)} \qquad (23)$$

which will also tend to the Shannon entropy at the limit of $q \to 1$. We hope that Eq. (23) will be useful for the future study in nonextensive statistical mechanics. Another question is that we get Eq. (23) based on the standard probability normalization, i.e., $\sum_i p_i = 1$. If one choose other normalization conditions, the different results will be derived.

## 4. Conclusions

In summary, with the help of variational uncertainty measurement $dI = d\overline{x} - \overline{dx}$, we generalized the definition of expectation value for different kinds of statistical methods including extensive and nonextensive. By this generalized uncertainty measurement, several maximizable entropy forms such as the Shannon entropy, Tsallis entropy and incomplete entropy have been derived directly. It's worth to point out that this generalized uncertainty measurement is valid for arbitrary dynamical system as long as the probability distribution function is given. It's proved that the uncertainty measurement of the system is nothing but the maximizable entropy. From experimental physics and other data sources of our real world, one can get many different probability distributions. For these distributions, which statistical method is suitable for? Or in other words, how to choose the concrete form of $g(p_i)$ to determine the entropy of the system? It's still an open question.

Acknowledgements

This work has been supported by the region des Pays de la Loire of France under grant number 2007-6088, by the National Natural Science Foundation (No. 10875100), People's Republic of China, and by the Science Research Fund (No. 07BS105), Huaqiao University, People's Republic of China.